\documentclass[aps,prb,superscriptaddress,floatfix,twocolumn]{revtex4}
\usepackage{graphicx}
\usepackage{amsmath,amsfonts,amssymb}
\newcommand{\nairo}{Na$_2$IrO$_3$}
\newcommand{\liiro}{Li$_2$IrO$_3$}
\newcommand{\airo}{A$_2$IrO$_3$}

\begin{document}

\title{ Analysis of the optical conductivity for $A_2$IrO$_3$
($A$ = Na, Li) from first principles}
\date{\today}

\begin{abstract}

  We present results for the optical conductivity of {\nairo} within
  density functional theory by including spin-orbit (SO) and
  correlation effects (U) as implemented in GGA+SO+U. We identify the
  various interband transitions and show that the underlying
  quasi-molecular-orbital nature of the electronic structure in
  {\nairo} translates into distinct features in the optical
  conductivity. Most importantly, the parity of the quasi-molecular
  orbitals appears to be the main factor in determining
  strong and weak optical
  transitions. We also present optical conductivity calculations for {\liiro}
  and discuss the similarities and differences with {\nairo}.

\end{abstract}

\author{Ying Li}
\affiliation{Institut f\"ur Theoretische Physik, Goethe-Universit\"at Frankfurt,
Max-von-Laue-Strasse 1, 60438 Frankfurt am Main, Germany}
\author{Kateryna Foyevtsova}
\affiliation{Quantum Matter Institute,
University of British Columbia, Vancouver,
British Columbia V6T 1Z4, Canada}
\author{Harald O. Jeschke}
\affiliation{Institut f\"ur Theoretische Physik, Goethe-Universit\"at Frankfurt,
Max-von-Laue-Strasse 1, 60438 Frankfurt am Main, Germany}
\author{Roser Valent{\'\i}}
\affiliation{Institut f\"ur Theoretische Physik, Goethe-Universit\"at Frankfurt,
Max-von-Laue-Strasse 1, 60438 Frankfurt am Main, Germany}

\maketitle

The family of honeycomb iridates {\airo} (A = Na, Li) has recently
been a subject of intensive discussion due to its complex electronic
and magnetic behavior arising from an interplay of spin-orbit effects,
correlations and lattice geometry. These materials are insulators and
order antiferromagnetically at low temperatures~\cite{Singh2010}. While
{\nairo} shows a zigzag-like magnetic pattern~\cite{Choi2012}, a spiral
order with a small nonzero wave vector inside the first Brillouin zone
has been reported for {\liiro}~\cite{Coldea1,Kimchi2014}. Photoemission and
optical conductivity measurements~\cite{Comin2012}
 for {\nairo} confirm this insulating
behavior with a gap of 340~meV. Attempts
to understand the different behavior of the end members through the
series (Na$_{1-x}$Li$_x$)$_2$IrO$_3$ have been
pursued~\cite{Manni2014,Cao2013} with partly contradicting results,
while Manni {\it et al.}~\cite{Manni2014} find that only for $x \le
0.25$ does the system form uniform solid solutions and otherwise the
system shows a miscibility gap and phase separates, Cao {\it et
 al.}~\cite{Cao2013} report a homogeneous phase at $x \sim 0.7$ with
a disappearance of long range magnetic order.

From a theoretical point of view, these materials have been
suggested to be a realization of the Heisenberg-Kitaev model with
bond-dependent anisotropic interactions between $j_{\rm eff} =1/2$
spin-orbit-coupled Ir
moments~\cite{Chaloupka2010,Imada2014,Katukuri2014,Rau2014,Reuther2014,Perkins2014}.
Such a model is obtained under the assumption of large spin-orbit
coupling, so that Ir 5d $t_{2g}$ orbitals can be written in terms
 of $j_{\rm eff} = 1/2$ and lower lying $j_{\rm eff} = 3/2$ relativistic
orbitals.

Alternatively, a description of the electronic structure of these
systems in terms of quasi-molecular orbitals was also recently
proposed~\cite{Mazin2012,Foyevtsova2013}. Following the observation that the
contributing energy scales in these systems are of the same order of
magnitude, namely the bandwidth for 5$d$ orbitals is 1.5-2~eV, the
onsite Coulomb repulsion U is about 1-2~eV, the Hund's coupling
constant is about 0.5~eV and the spin-orbit coupling is $\lambda \sim$
0.4-0.5~eV it was shown~\cite{Mazin2012,Foyevtsova2013}
 that the underlying electronic
behavior can be described in terms of molecular orbitals formed by the
Ir $t_{2g}$ states on an hexagon with each of the three $t_{2g}$ Ir
orbitals on a site contributing to three neighboring molecular
orbitals. In Ref.~\onlinecite{Foyevtsova2013} it was further
demonstrated that both
descriptions, i.e. a localized description in terms of $j_{\rm eff}$
and an itinerant description in terms of molecular orbitals are
equally compatible.

Presently,
only optical conductivity $\sigma (\omega)$ measurements
for {\nairo} are available~\cite{Comin2012,Sohn2013}.
$\sigma (\omega)$ in {\nairo} shows a broad peak structure at 1.5~eV~\cite{Comin2012} (1.66~eV in Ref.~\onlinecite{Sohn2013})
and smaller peak structures at 0.52~eV, 0.72~eV, 1.32~eV, 1.98~eV~\cite{Sohn2013}. These features have
been interpreted in terms of dominant $j_{\rm eff}$= 3/2 and $j_{\rm
 eff}$= 1/2 transitions~\cite{Sohn2013}.
With the aim of further unveiling the origin of different
behavior in {\nairo} and
{\liiro}, we revisit the optical conductivity in
{\nairo} with density functional theory calculations
and show that the nature
of the various
interband transitions observed experimentally
can be understood in terms of the parity of the underlying
molecular orbital description.
 In contrast, the optical conductivity behavior that we predict
for {\liiro} shows an increase in weight at low
energies with respect to {\nairo}
due to a strong mixing of quasi-molecular orbital character,
absent in {\nairo}.

For our density functional theory (DFT) calculations we use
the full-potential linearized augmented plane-wave (LAPW) method as
implemented in the code WIEN2k~\cite{Wien2k}. The PBE generalized
gradient approximation (GGA)~\cite{Perdew1996} was employed as exchange
correlation functional and the basis-size controlling parameter
$RK_{max}$ was set to 8. A mesh of 450 ${\bf k}$ points in the first
Brillouin zone (FBZ) for the self-consistency cycle was used. In
order to have a good description of the experimentally observed
optical gap in {\nairo}, magnetism as well as a $U_{\rm eff}=2.4$~eV
as implemented in GGA+U~\cite{Anisimov1993} had to be included
in the calculations.
 Relativistic effects were taken into account within the second variational
approximation (GGA+SO+U). For the optical properties, we employed the
optics code package~\cite{Draxl} in WIEN2k. The optical properties
were calculated with 1568 ${\bf k}$ points in the FBZ.

\begin{figure}
\includegraphics[angle=0,width=0.45\textwidth]{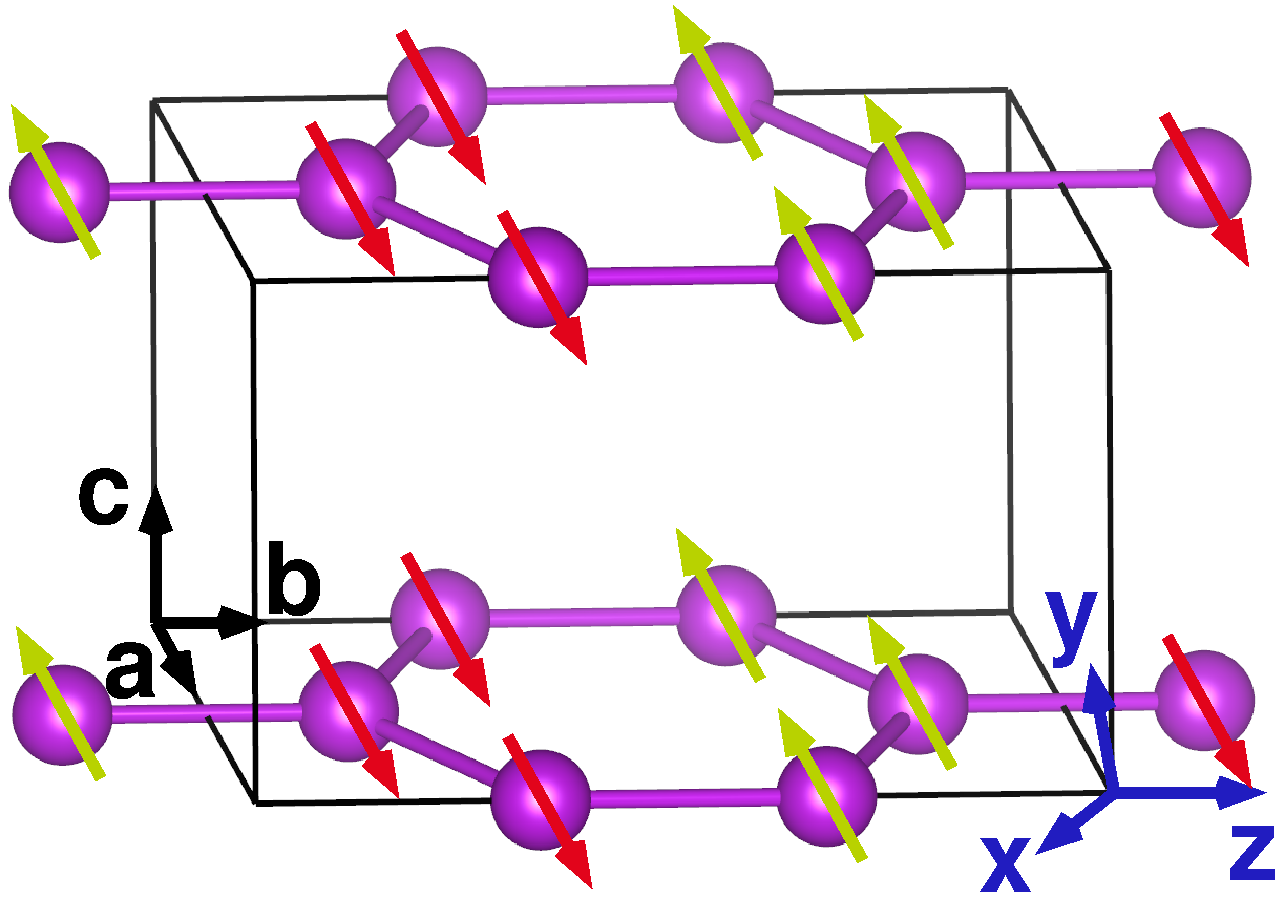}
\caption{ (Color online) Ir honeycomb layers of {\nairo}. The black
 axis ${\bf a}$, ${\bf b}$, and ${\bf c}$ are the vectors of the unit
 cell, while the dark blue axes ${\bf x}$, ${\bf y}$, ${\bf z}$ are
 the Cartesian axes. The red and green arrows show the zigzag AFM
 phase.}
\label{structure}
\end{figure}

 The imaginary part of the interband contribution to the dielectric
 function is given by~\cite{Draxl,Ferber2010}:
 \begin{equation}\begin{split}
{\rm Im}\epsilon _{\alpha \beta }(\omega )\propto\frac{1}{\omega ^{2}}\sum_{c,v}\int d{\bf k}&\left\langle c_{\bf k}|p^{\alpha }|v_{\bf k}\right\rangle\left\langle v_{\bf k}|p^{\beta }|c_{\bf k}\right\rangle\\
&\times\delta (\varepsilon
_{c_{\bf k}}-\varepsilon _{v_{\bf k}}-\omega ).
\end{split}\end{equation}
Here, $\alpha$ and $\beta$ indicate directional components, $p$ is the momentum
operator, and $\omega $ corresponds to the energy of the
photon. $c_{\bf k}$ denotes a state in the conduction band with the
energy $\varepsilon _{c_{\bf k}}$ and $v_{\bf k}$ is a state in the
valence band with the energy $\varepsilon _{v_{\bf k}}$. By absorbing
photon energy, the electrons transit from $v_{\bf k}$ to $c_{\bf
 k}$. The real part of the dielectric function can be evaluated from
the imaginary part using the Kramers-Kronig relation. In this work we focus on the analysis of the real part of the optical
conductivity
\begin{equation}
{\rm Re}\,\sigma _{\alpha \beta }(\omega )=\frac{\omega }{4\pi }{\rm Im}\,\epsilon _{\alpha
\beta }(\omega ).
\end{equation}
For our DFT analysis we used the experimental structure of
{\nairo} given in Ref.~\onlinecite{Choi2012} which agrees well with the
relaxed structure~\cite{Choi2012,Manni2014}
and performed GGA+SO+U ($U=3$~eV, $J_{\rm H}=0.6$~eV, $U_{\rm eff} =
U-J_{\rm H}=2.4$~eV) calculations in the zigzag antiferromagnetic
(AFM) ordered phase (see Fig.~\ref{structure}) with the magnetization
parallel to the ${\bf a}$ direction~\cite{Liu2011}.

\begin{figure}[tbp]
 \includegraphics[angle=0,width=0.48\textwidth]{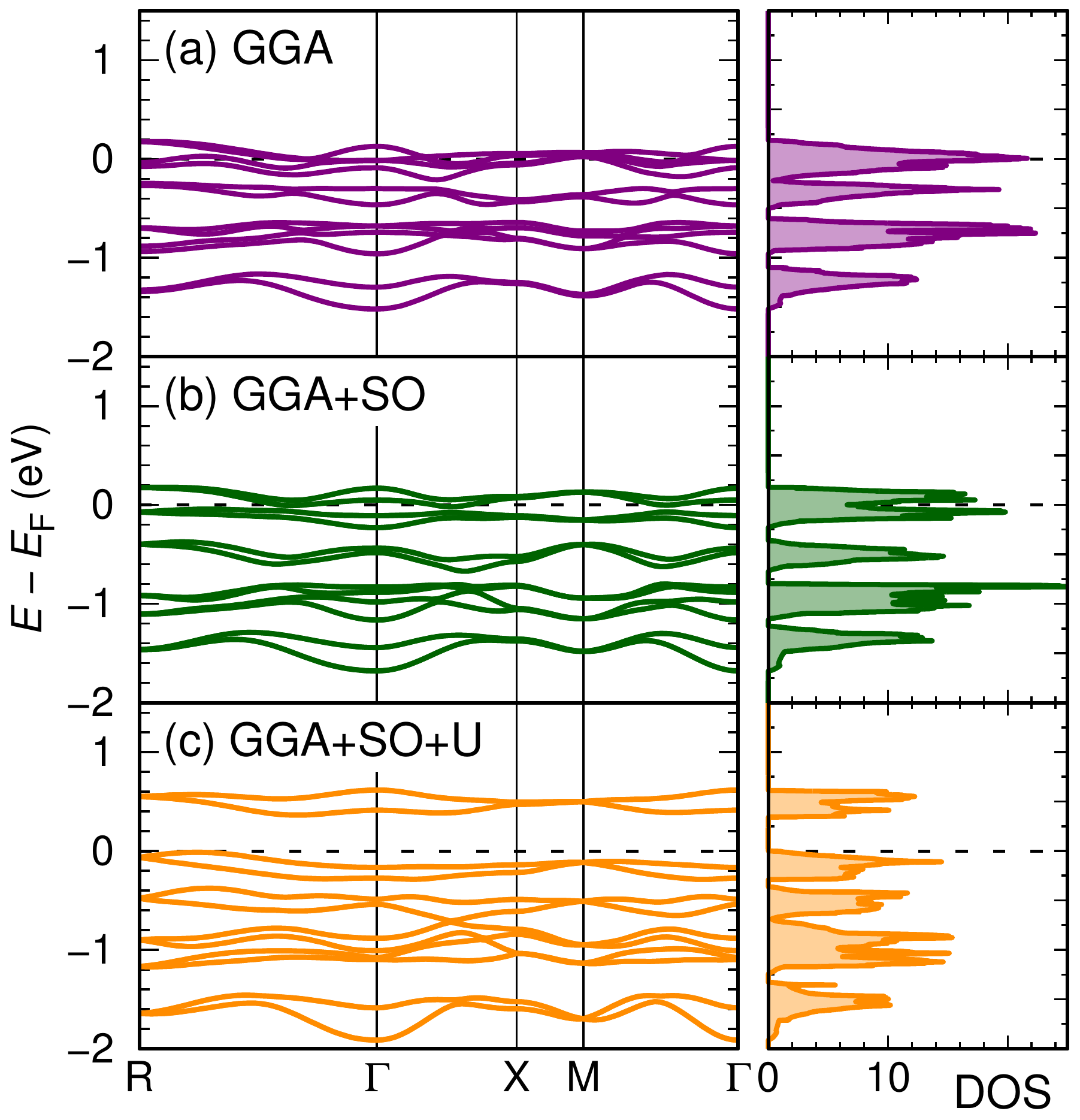}
 \caption{(Color online) Ir 5$d$ $t_{2g}$ DOS and band structures
for {\nairo}, obtained with (a) GGA, (b) GGA+SO, and (c) GGA+SO+U (U=3~eV, J$_{\rm H}$=0.6~eV, U$_{\rm eff}$ = U-J$_{\rm H}$=2.4~eV).}
\label{band-dos}
\end{figure}

The density of states (DOS) and band structures for
{\nairo} within GGA, GGA+SO, and GGA+SO+U are shown in
Fig.~\ref{band-dos}. Compared to the non-relativistic GGA DOS, a
suppression of the DOS at $E_{\text{F}}$ is clearly visible in the
relativistic GGA+SO calculation. In GGA+SO+U, a 341~meV gap can be
obtained as reported experimentally~\cite{Comin2012}. Note
that in the zigzag AFM phase there are 4 iridium atoms
per unit cell and therefore
the number of bands doubles to 12 $t_{2g}$ in Fig.~\ref{band-dos}.

The monoclinic symmetry allows for
four independent components of the optical conductivity tensor defined
as $\sigma _{xx}$, $\sigma _{yy}$, $\sigma _{zz}$, $\sigma _{xy}$
\begin{equation}%
\left(\begin{smallmatrix} J_x\\ J_y\\ J_z \end{smallmatrix}\right)%
=\left(\begin{smallmatrix} {\sigma}_{xx}&{\sigma}_{xy}&0\\{\sigma}_{xy}&{\sigma}_{yy}&0\\0&0&{\sigma_{zz}} \end{smallmatrix}\right)
\left(\begin{smallmatrix} E_x\\ E_y\\ E_z \end{smallmatrix}\right).
\end{equation}%
 The Cartesian directions are
shown in Fig.~\ref{structure}. ${\bf z}$ is parallel to the
${\bf b}$ direction and lies in the Ir hexagonal plane,
 while ${\bf x}$ and ${\bf y}$ are in the $ac$
plane. Spin-orbit coupling also induces small non-zero contributions
to the $\sigma_{xz}$ and $\sigma _{yz}$ components.

\begin{figure}
\includegraphics[angle=0,width=0.5\textwidth]{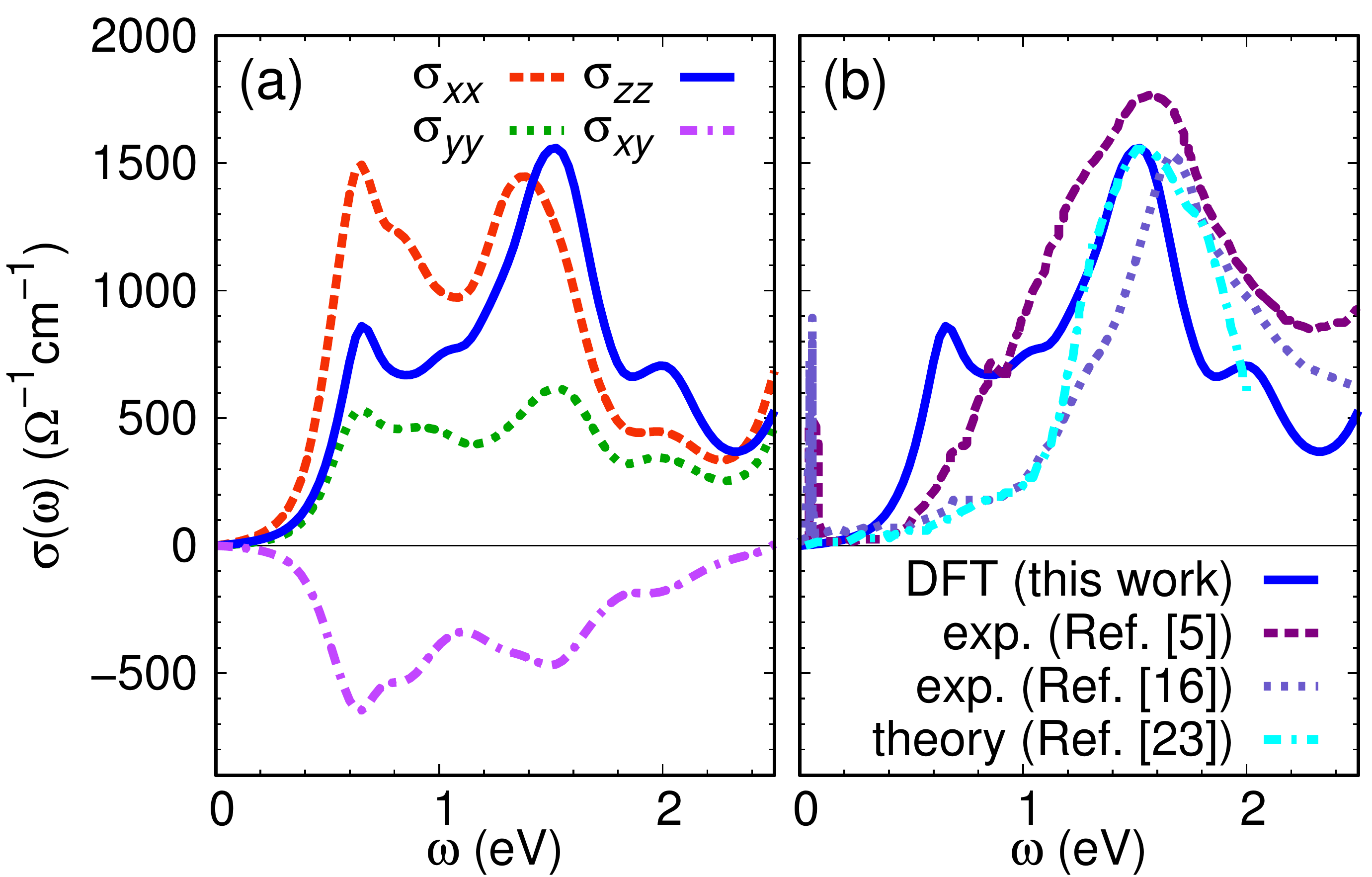}
\caption{ (Color online) (a) Optical conductivity tensor components
 $\sigma _{xx}$, $\sigma _{yy}$, $\sigma _{zz}$, $\sigma _{xy}$
 and (b) DFT $\protect\sigma _{zz}$ for {\nairo} compared with experiment %
 \protect~\cite{Comin2012,Sohn2013} and theory data \protect~\cite{Kim2014}.}
\label{optxyz}
\end{figure}
In Fig.~\ref{optxyz}~(a), we present the calculated four dominant optical
conductivity tensor components for {\nairo} in the low-frequency region.
Only $\sigma _{zz}$ corresponds to the in-plane
optical conductivity and in Fig.~\ref{optxyz}~(b) we compare
this component with the experimental
results~\cite{Comin2012,Sohn2013} as well as with a 4-site iridium cluster
calculations by Kim {\it et al}~\cite{Kim2014}. Both,
 our DFT calculations and the cluster calculations~\cite{Kim2014}
 show the presence of
a dominant peak at $\omega=$ 1.5~eV as observed in experiment.
However, the DFT results have a richer structure and capture
the multi-peak behavior of the experimental observations.

\begin{figure}
\includegraphics[angle=0,width=0.5\textwidth]{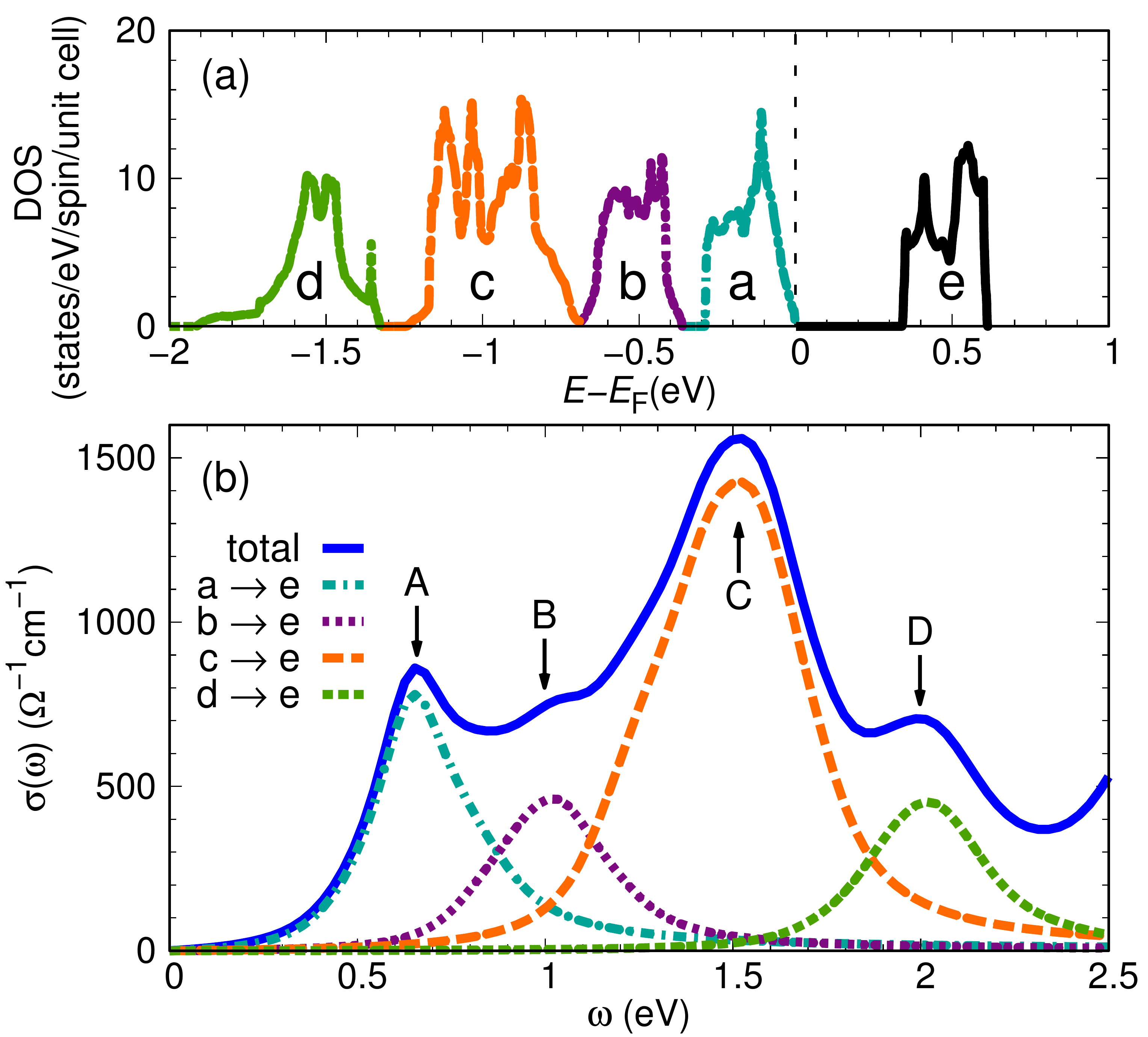}
\caption{(Color online) GGA+SO+U density of states (a)
and contributions from different {\it d}-{\it d} transitions (b). The a, b, c, d, e label the 5 states.}
\label{optdis}
\end{figure}

In order to disentangle the origin of the various features present in
the optical conductivity in Fig.~\ref{optxyz}~(b),
we display in Fig.~\ref{optdis} the various interband
processes. For that purpose, we label
in Fig.~\ref{optdis}~(a) the valence states $v_s$
 as a, b, c, d and the
conduction states $c_s$ as e.
We identify four peaks in $\sigma(\omega)$ (Fig.~\ref{optdis}~(b));
 peaks A, B, C, D
correspond to the transitions from a, b, c, d to e states,
respectively. The analysis of the electronic structure in terms of
quasi-molecular orbitals~\cite{Mazin2012,Foyevtsova2013} predicts a
clear odd/even parity
related to the symmetry of the quasi-molecular orbitals, i.e.
odd $B_{1u}$, even $E_{1g}$, odd $E_{2u}$ and even $A_{1g}$.
Even though the zigzag magnetic order used for the calculations
mixes states
of different parities, we find in our analysis of the magnetic
quasi-molecular orbitals that the dominating
parity contribution to a given state matches
the parity of this state's counterpart in the paramagnetic phase.
This, in particular, allows us to compare our spin-polarized
calculations with the measurements performed above the magnetic
transition temperature.

In the GGA+SO+U calculations we find that the states
a, b, c, d, and e are predominantly of
even, odd, even, odd, and odd parity, respectively;
note that in the presence of spin-orbit coupling the states from the
upper triad cannot be identified in terms of quasi-molecular orbitals, however we can still discern the dominant parity.
Since the dielectric tensor matrix elements involved in the
optical interband transitions are of the form
$\left\langle v_s|{\bf E}\cdot{\bf r}|c_s\right\rangle$ with ${\bf E}\cdot{\bf r}$ being
an odd parity operator, clearly, transitions between states of the
same parity will be strongly suppressed whereas transitions between
states of different parity will dominate. This is reflected in the
large peak at 1.5~eV (peak C) that corresponds to a predominantly even to
odd parity transition, followed by peak A (predominantly even to odd), while peaks B and D
are of transitions
between predominantly equal (odd) parity states and are strongly suppressed
(see Ref.~\onlinecite{Mazin2012}, Supplement). The optical
conductivity is therefore an important measure of the underlying
molecular orbital structure in {\nairo}.

\begin{figure}
\includegraphics[angle=0,width=0.5\textwidth]{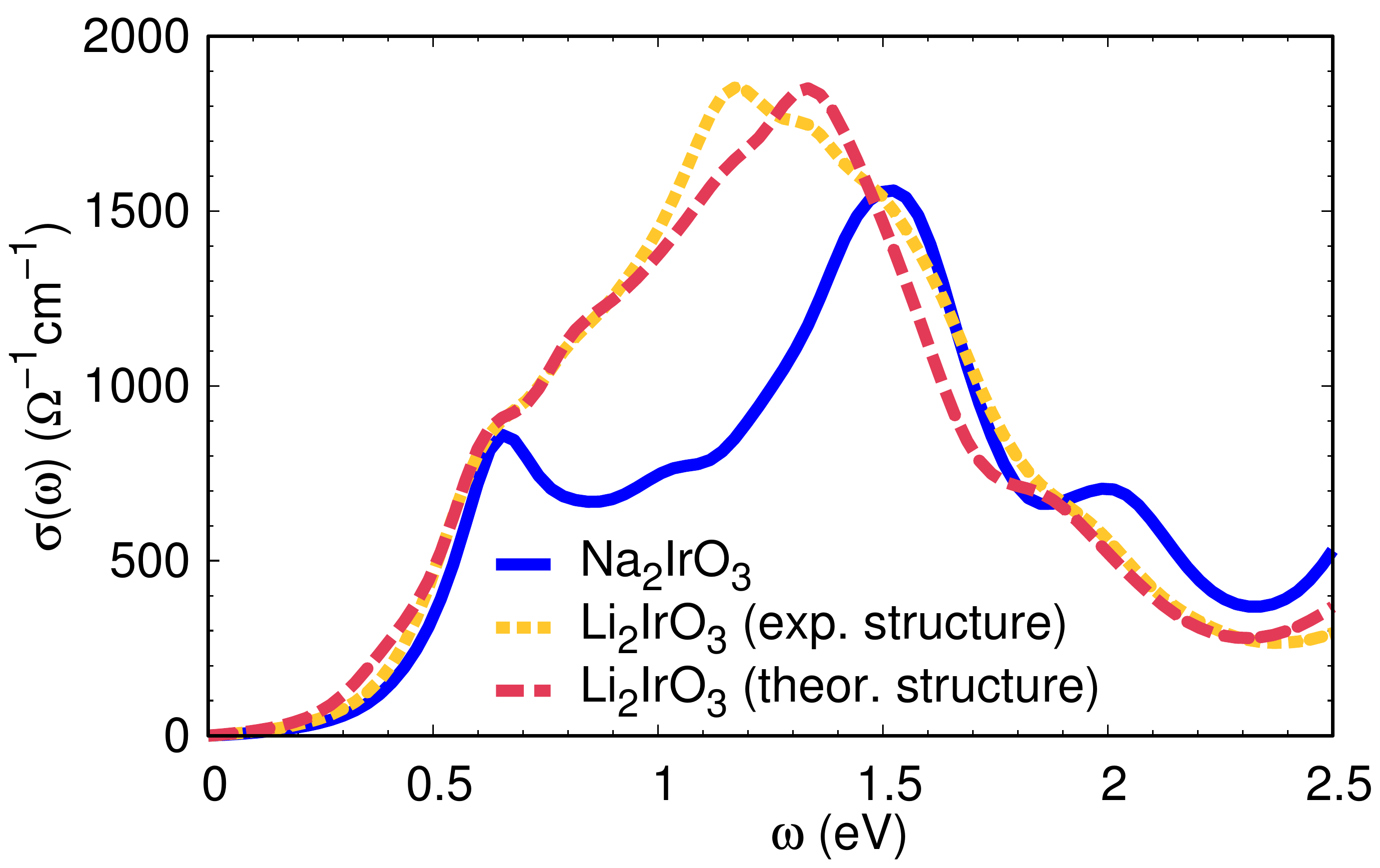}
\caption{(Color online) Comparison of the optical conductivity
 $\protect\sigma _{zz}$ between {\nairo} (experimental structure) and
 {\liiro} (experimental and theoretically predicted structure).}
\label{opt_compare}
\end{figure}

We now proceed with the calculation of the optical conductivity for
{\liiro}. The {\liiro} structure is known from powder x-ray
diffraction~\cite{Gretarsson2013} and from a careful DFT structure
prediction using a spin-polarized GGA+SO+U exchange correlation
functional~\cite{Manni2014}. As these two structures differ slightly
and also show small but significant differences in electronic
structure~\cite{supplement}, we determine the optical conductivity for
both of them. Experiments indicate that the gap of {\liiro} is of the same order of magnitude as in {\nairo}~\cite{Gretarsson2013} or a bit smaller~\cite{Jenderka}. We find that magnetism
and a $U_{\rm eff}=2.4$~eV are necessary to open a gap of about 318~meV for the experimental structure while $U_{\rm eff}=2.0$~eV is necessary to open a gap of about 307~meV for the theoretical structure.
Even though a spiral order has been suggested from experiment~\cite{Coldea1},
we have considered, for simplicity, a zigzag magnetic order
as in {\nairo} for the calculations.

\begin{figure}[htbp]
\includegraphics[angle=0,width=0.48\textwidth]{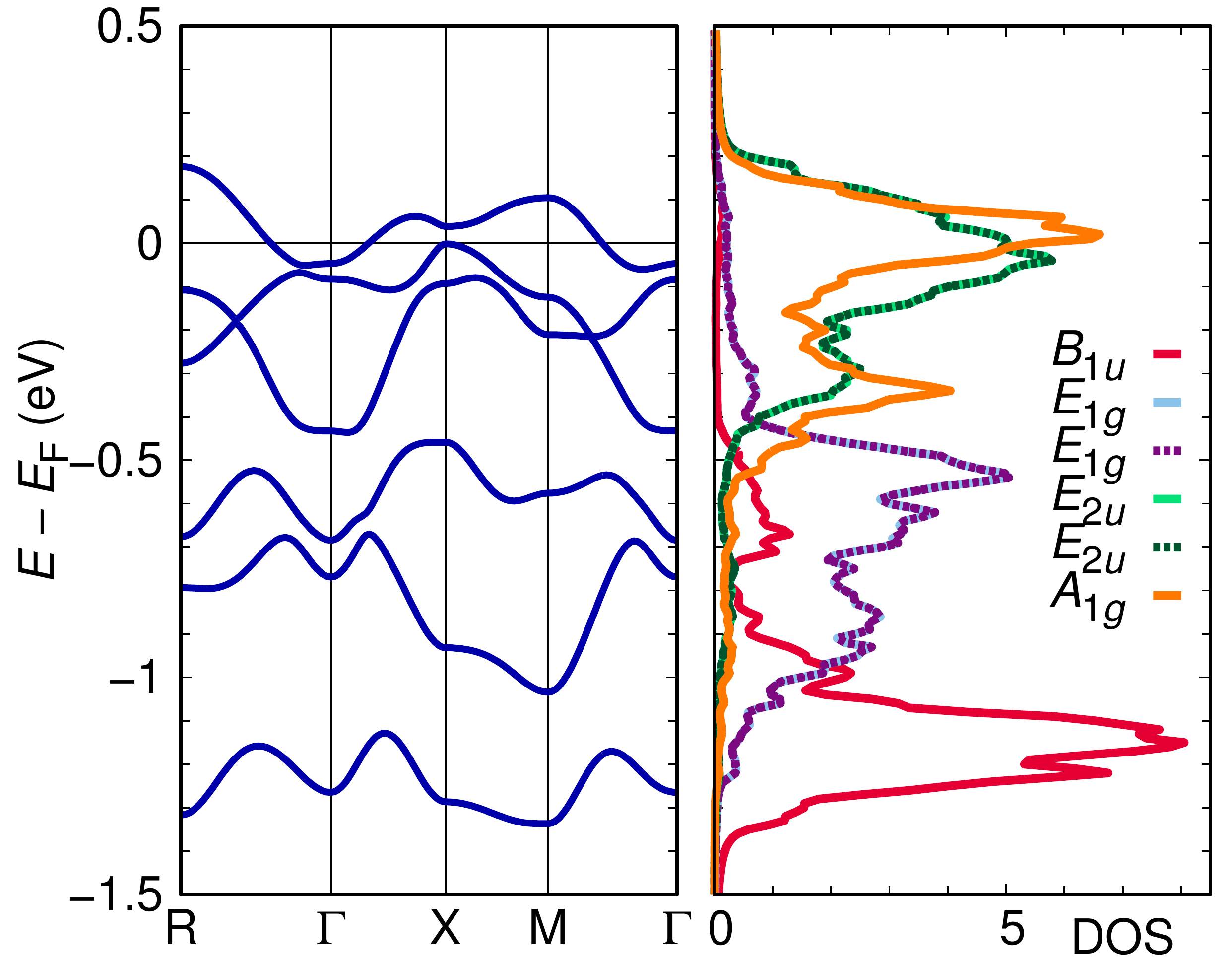}
\caption{(Color online) Nonrelativistic nonmagnetic band structure and
 density of states of the theoretically predicted {\liiro} structure,
 projected onto quasi-molecular orbitals.}
\label{qmo_lirelax}
\end{figure}

We compare the optical conductivities of {\nairo} and {\liiro} in
 Figure~\ref{opt_compare}. While the dominant peak in the {\nairo}
optical conductivity is at 1.5~eV, we find it at 1.17~eV for the
experimental structure and at 1.33~eV for the theoretical
structure of {\liiro}. Also, we observe an increase
of the optical conductivity weight between 0.66~eV
and 1.48~eV with respect to {\nairo}. In order to analyze this behavior,
 we project the nonmagnetic GGA electronic
structure of {\liiro} onto the quasi-molecular orbital basis
(see Fig.~\ref{qmo_lirelax}).
We observe that the separation of the density of
states into isolated narrow bands of unique quasi-molecular orbital
characters is much less clean than in
{\nairo}~\cite{Mazin2012,Foyevtsova2013} and resembles
the case of Li$_2$RhO$_3$~\cite{Mazin2013}. In {\liiro},
there is overlapping between $B_{1u}$ and $E_{1g}$ states and between
$E_{1g}$ and $A_{1g}$/$E_{2u}$ states as shown in
Fig.~\ref{qmo_lirelax}. This strong mixing of character, which
remains in the magnetic calculations, explains why
the B peak in {\liiro} is much stronger than in {\nairo}; the suppressed
odd to odd transition in {\nairo} evolves into a mixture
of enhanced and
suppressed transitions in {\liiro}.

We would like to emphasize that all above DFT calculations
have been performed with inclusion of spin-orbit effects, and,
strictly speaking, neither the $t_{2g}$ (or the linear combination
of $t_{2g}$ states forming quasi-molecular orbitals) nor spin are
well defined entities.
Nevertheless, we have shown that
the main features observed
in optical conductivity are related to the underlying symmetries of
the molecular orbital basis, which is a manifestation of the fact
that spin-orbit coupling is not the only determining interaction
in these materials.

In summary, we have investigated the optical conductivity in
{\nairo} and {\liiro} by performing magnetic GGA+SO+U calculations.
Magnetism and a nonzero $U$ were necessary in order to reproduce
the experimental insulating gap in both systems.
Using the fact that the narrow bands
of {\nairo} are well described in terms of quasi-molecular orbitals, we
showed that the strength of the various interband contributions
 to the optical conductivity can be well described in
 terms of the
parity of the quasi-molecular orbitals, namely weight suppression in
 like-parity
transitions and weight enhancement
in unlike-parity transitions. We also predict the shape
of the optical conductivity for {\liiro}. Contrary to {\nairo},
in {\liiro} the quasi-molecular
orbitals strongly overlap and parities mix. This explains
 the relative weight differences in the optical conductivity between
{\liiro} and {\nairo}.

\begin{acknowledgments} We would like to thank J. Orenstein, R. Coldea,
 J. Analytis, X. Xi, I. I. Mazin and G. Khaliullin
 for very useful discussions.
Y.L. acknowledges support through a China
 Scholarship Council (CSC) Fellowship. H.O.J and R.V. acknowledge
 support by the Deutsche Forschungsgemeinschaft through grant SFB/TR
 49.
\end{acknowledgments}

\clearpage
\renewcommand{\thetable}{S\Roman{table}}
\renewcommand{\thefigure}{S\arabic{figure}}
\renewcommand{\thesection}{}
\renewcommand{\thesubsection}{S\arabic{subsection}}
\setcounter{figure}{0}
\setcounter{table}{0}

\section*{Supplementary Information}
\subsection{The effect of broadening}

\begin{figure}[htbp]
  \includegraphics[angle=0,width=0.48\textwidth]{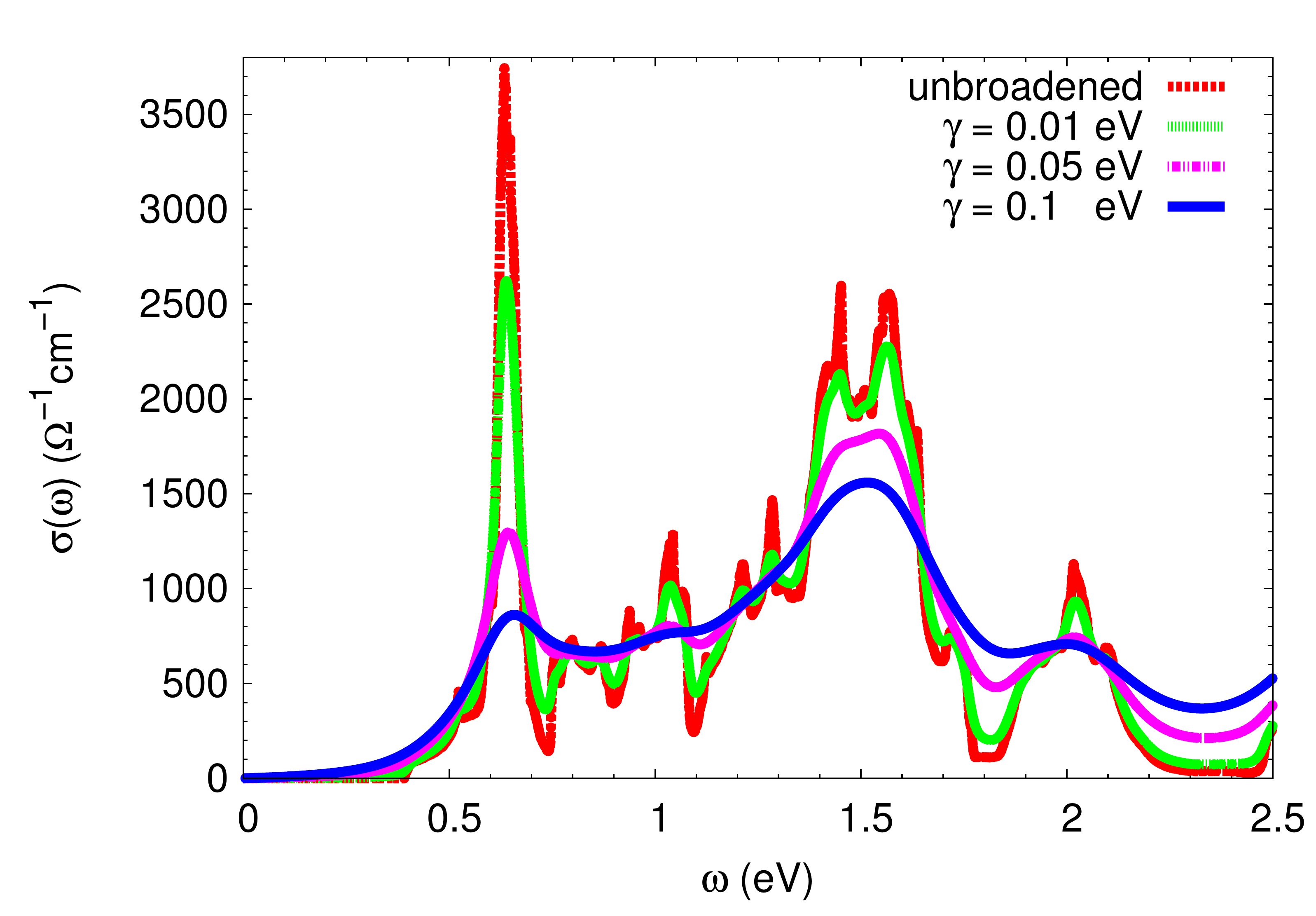}
  \caption{(Color online) Optical conductivity for {\nairo} without and with different broadenings $\gamma= 0.01$, 0.05, and 0.1~eV.}
\label{broad}
\end{figure}

The unbroadened theoretical spectra require a substantial broadening
for a meaningful comparison with the experimantal data.  We use
different values of the Lorentzian broadening $\gamma$ within the
WIEN2k optics package to obtain the broadened theoretical spectra in
Fig.~\ref{broad}. We choose $\gamma=0.1$~eV as the broadening to
compare with experiment.

\subsection{The electronic structure for {\liiro}}

\begin{figure*}
  \includegraphics[angle=0,width=0.49\textwidth]{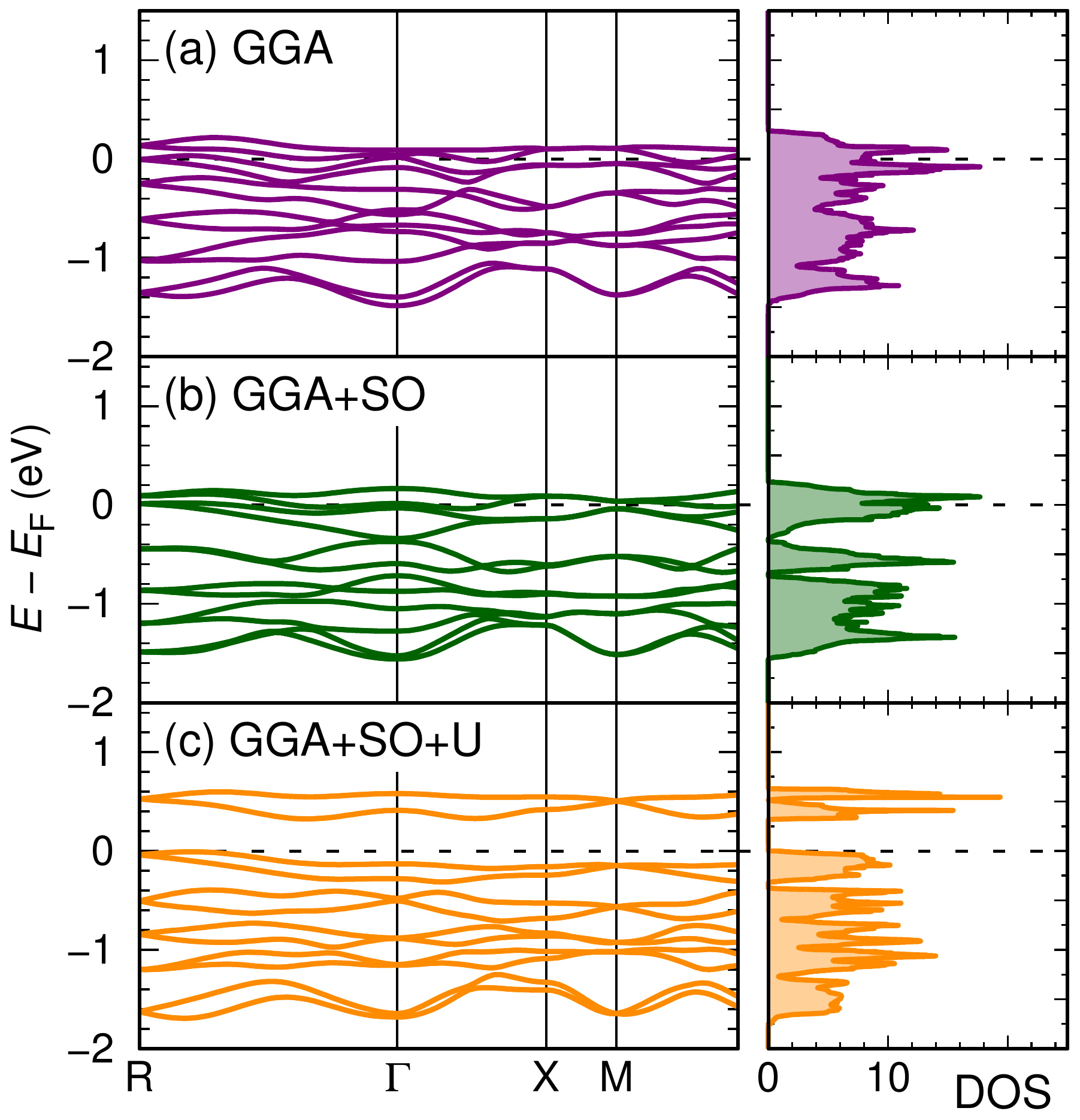}%
  \includegraphics[angle=0,width=0.49\textwidth]{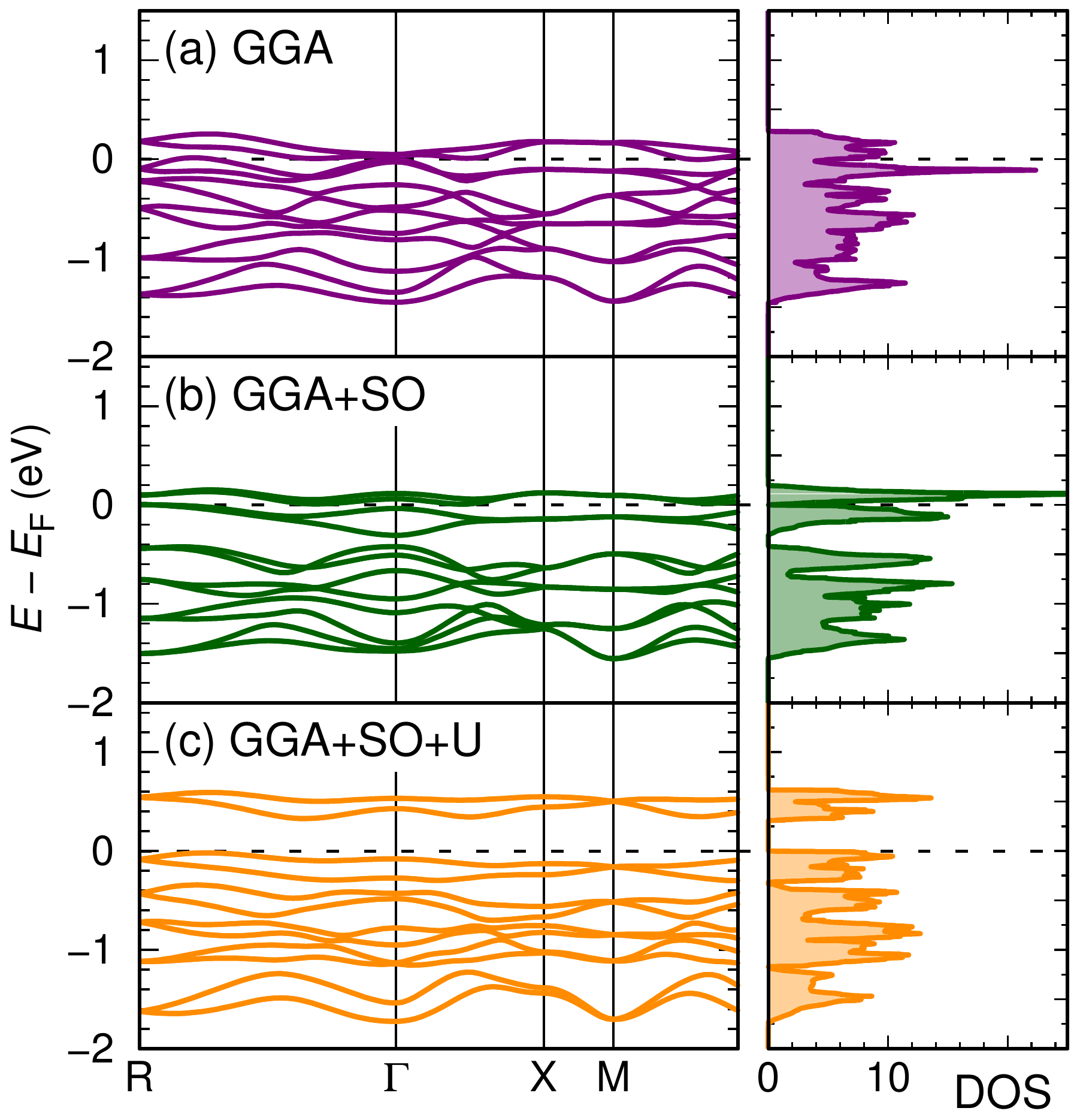}
  \caption{(Color online) Ir 5$d$ $t_{2g}$ density of states and band structures for experimental Li$_2$IrO$_3$ structure (left) and for the theoretical Li$_2$IrO$_3$ structure (right), obtained with (a) GGA, (b) GGA+SO, and (c) GGA+SO+U.}
\label{band-dos-lire-exp}
\end{figure*}

For performing DFT calculations on {\liiro}, we used the
experimentally determined atomic positions and lattice parameters, the
latter being $a=5.172$~{\AA}, $b=8.926$~{\AA}, $c=5.122$~{\AA},
$\alpha=\gamma =90^{\circ }$, $ \beta =109.91^{\circ }$ (see
Ref.~S\onlinecite{Gretarsson2013sup}). The corresponding electronic
structure is shown in Figure~\ref{band-dos-lire-exp} (left). We also
used the crystal structure predicted in Ref.~S\onlinecite{Manni2014sup}
with spin polarized GGA+SO+U. In this case, the lattice constants are
$a=5.21518$~{\AA}, $b=9.01171$~{\AA}, $c=5.14869$~{\AA},
$\alpha=\gamma =90^{\circ }$, $ \beta =109.89^{\circ }$ and the
electronic structure is presented in Figure~\ref{band-dos-lire-exp}
(right).
The spin-polarized GGA+SO calculation for the theoretical structure
converges to a nonmagnetic solution and the DOS is similar to that of
{\nairo}, which has a very low DOS at $E_{\text{F}}$.
\vspace{0.5cm}

\subsection{Optical conductivity contributed by different {\it d}-{\it d}
transitions for {\liiro}}
\begin{figure*}[htbp]
\centering
\includegraphics[angle=0,width=0.49\textwidth]{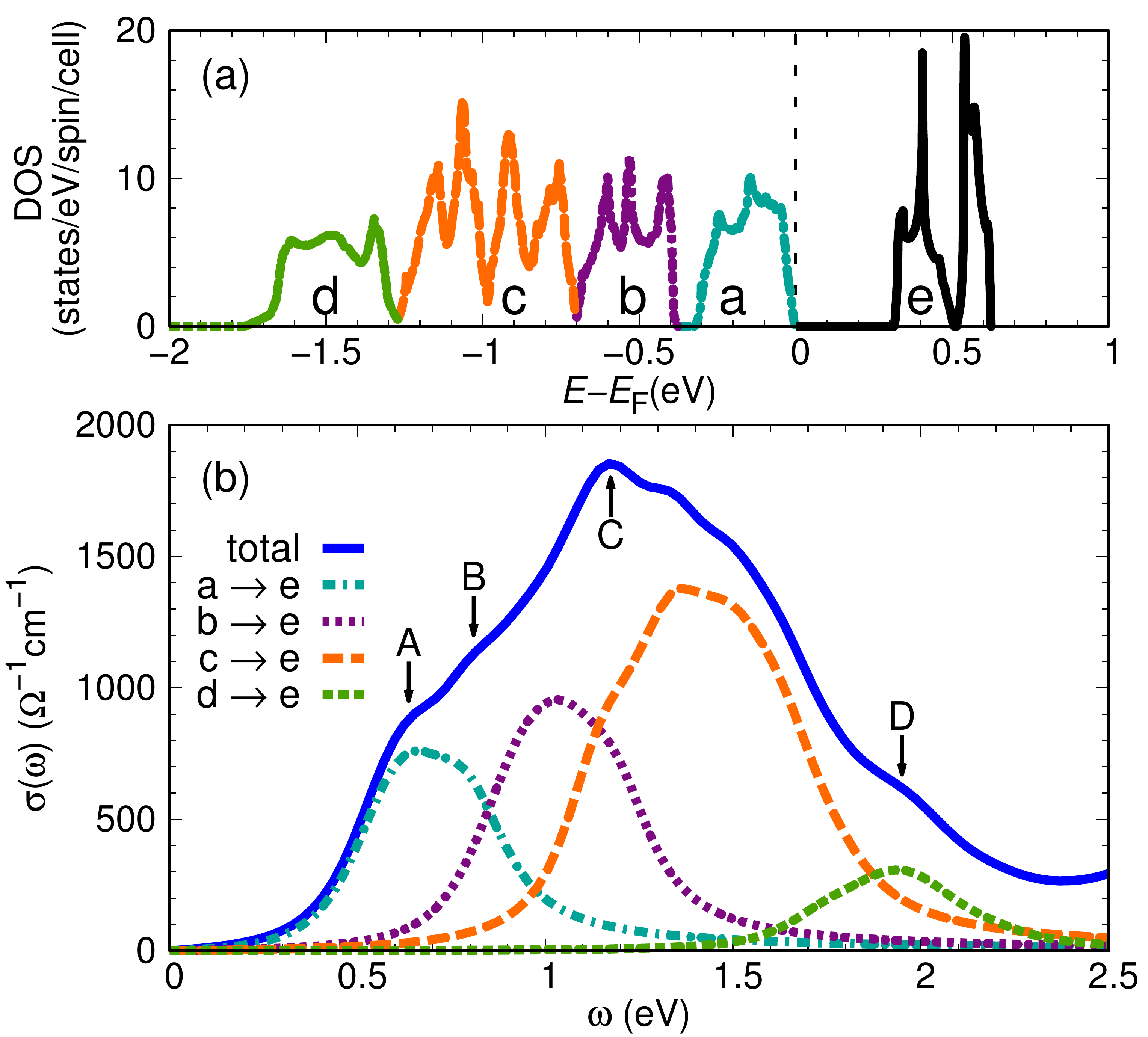}%
\includegraphics[angle=0,width=0.49\textwidth]{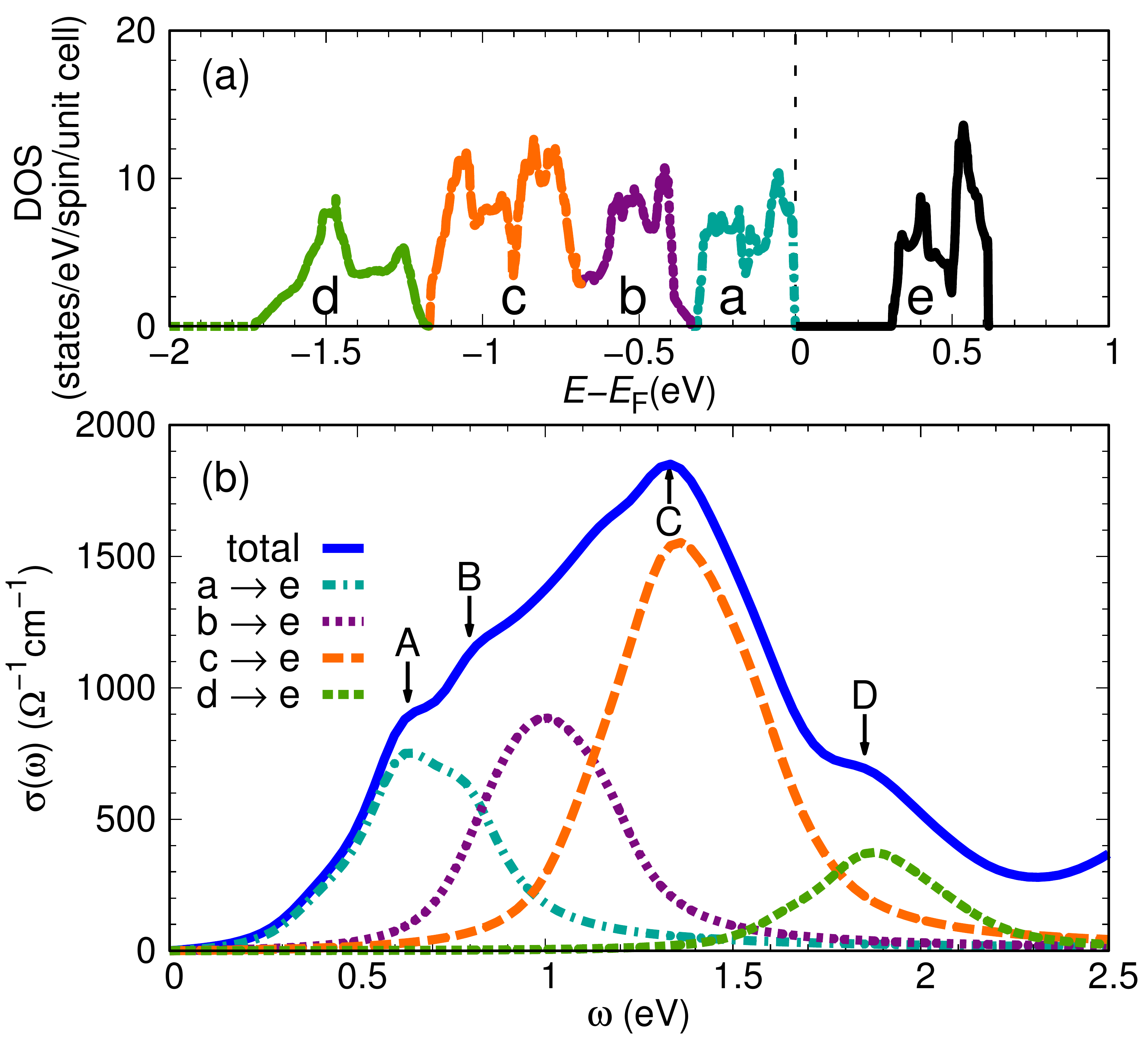}
\caption{ (Color online) Density of states (a) and contributions from
  different {\it d}-{\it d} transitions (b) for the experimental
  Li$_2$IrO$_3$ structure (left) and for the theoretical Li$_2$IrO$_3$
  structure (right). The a, b, c, d, e label the 5 states.}
\label{optdis_exp}
\end{figure*}
We display the optical conductivity contributions from different {\it
  d}-{\it d} transitions for both the experimental and theoretical
crystal structures of {\liiro} in
Figure~\ref{optdis_exp}.
In {\liiro} the transition from the b state to the e state is strongly enhanced
compared to {\nairo}.  In {\nairo} this transition was
between bands of dominantly the same  parity
 and therefore relatively strongly suppressed. In {\liiro}, due
to the quasi-molecular orbital overlap and parity mixing,
 the B peak even dominates
over the A peak. The experimental structure shows similar features.

\renewcommand*{\citenumfont}[1]{S#1}
\renewcommand*{\bibnumfmt}[1]{[S#1]}

\end{document}